%% file: conference_101719.tex
\documentclass[conference]{IEEEtran}
\IEEEoverridecommandlockouts
\usepackage{cite}
\usepackage{amsmath,amssymb,amsfonts}
\usepackage{algorithmic}
\usepackage{graphicx}
\usepackage{subfig}
\usepackage{textcomp}
\usepackage{xcolor}
\def\BibTeX{{\rm B\kern-.05em{\sc i\kern-.025em b}\kern-.08em
    T\kern-.1667em\lower.7ex\hbox{E}\kern-.125emX}}

\usepackage{hyperref}
\usepackage{bm}
\usepackage[group-separator={,}]{siunitx}
\usepackage{bbm}
\usepackage{bbold}
\usepackage{multirow}

\usepackage{tabularx}
\usepackage{adjustbox}
\usepackage{xspace}

\usepackage{xcolor}

\usepackage[ruled,linesnumbered]{algorithm2e}
\SetKw{KwWith}{with}
\SetKw{KwDownto}{downto}
\SetKwFor{ForEach}{for each}{do}{end}
\SetKwInput{Input}{Input}
\SetKwInOut{Output}{Output}
\SetKwProg{Function}{Function}{:}{end}
\DontPrintSemicolon

\def\method{\emph{SR-PredAO}\xspace}

\newcommand{\splitatcommas}[1]{%
  \begingroup
  \begingroup\lccode`~=`, \lowercase{\endgroup
    \edef~{\mathchar\the\mathcode`, \penalty0 \noexpand\hspace{0pt plus 1em}}%
  }\mathcode`,="8000 #1%
  \endgroup
}

\newtheorem{lemma}{Lemma}[section]
\def\done{\hspace*{\fill} $ \framebox[2mm]{}$ \medskip }

\AtBeginDocument{%
  \providecommand\BibTeX{{%
    \normalfont B\kern-0.5em{\scshape i\kern-0.25em b}\kern-0.8em\TeX}}}

\def\BibTeX{{\rm B\kern-.05em{\sc i\kern-.025em b}\kern-.08em
    T\kern-.1667em\lower.7ex\hbox{E}\kern-.125emX}}

\DeclareMathAlphabet{\mathbb}{U}{msb}{m}{n}
\usepackage{booktabs}

\usepackage{fancyhdr}

\begin{document}

\title{SR-PredictAO: Session-based Recommendation with High-Capability Predictor Add-On}

\author{
\IEEEauthorblockN{Ruida WANG, Raymond Chi-Wing WONG, Weile TAN}
\IEEEauthorblockA{\textit{The Hong Kong University of Science and Technology} \\
Kowloon, Hong Kong \\
rwangbr@connect.ust.hk, raywong@cse.ust.hk, wtanae@connect.ust.hk}
}

\maketitle

\input{0-abstract}


\input{1-introduction}
\input{2-related-work}
\input{3-preliminary.tex}

\input{4-model.tex}
\input{5-experiments.tex}
\input{6-conclusion.tex}

\bibliographystyle{IEEEtran}
\bibliography{references}

\input{7-appenidx.tex}

\end{document}

%% file: 0-abstract.tex
\begin{abstract}

    Session-based recommendation, aiming at making the prediction of the user's next item click based on the information in a single session only, even in the presence of some random user's behavior, is a complex problem. This complex problem requires a high-capability model of predicting the user's next action. Most (if not all) existing models follow the encoder-predictor paradigm where all studies focus on how to optimize the encoder module extensively in the paradigm, but they overlook how to optimize the predictor module. In this paper, we discover the critical issue of the low-capability predictor module among existing models. Motivated by this, we propose a novel framework called \emph{\underline{S}ession-based \underline{R}ecommendation with \underline{Pred}ictor \underline{A}dd-\underline{O}n} (SR-PredictAO). In this framework, we propose a high-capability predictor module which could  alleviate the effect of random user's behavior for prediction. It is worth mentioning that this framework could be applied to any existing models, which could give opportunities for further optimizing the framework. Extensive experiments on two real-world benchmark datasets for three state-of-the-art models show that \emph{SR-PredictAO} out-performs the current state-of-the-art model by up to 2.9\% in HR@20 and 2.3\% in MRR@20. More importantly, the improvement is consistent across almost all the existing models on all datasets, and is statistically significant, which could be regarded as a significant contribution in the field.

\end{abstract}

%% file: 1-introduction.tex
\section{Introduction}\label{intro}
Next-item recommender systems show their importance in the current age of e-commerce by accurately predicting the user's subsequent behavior. \textit{Session-based recommendation} is one recent hot topic of the next-item recommender. It is different from the \emph{general next-item recommendation systems}, which put great attention on a specific group of existing users with a large number of historical behavior records to perform the next-item prediction. The \emph{session-based recommendation}, as its name indicates, groups all the activities in the basic unit of the session and is based only on the information within a single session. The idea of session-based recommendation systems comes from \cite{youtube}. It shows that intra-session-dependencies have a more significant impact than inter-session dependencies on the user's final decision to view the next item. In particular, the user's next-item behavior is usually related to behaviors in the current session. For example, a user's behavior in buying phone accessories in one session has a relatively low connection to his/her action of buying clothes two days ago but has a strong relationship with his/her visit to a phone charger in the same session.

\begin{figure*}[t]
    \centering
    \includegraphics[width=\linewidth]{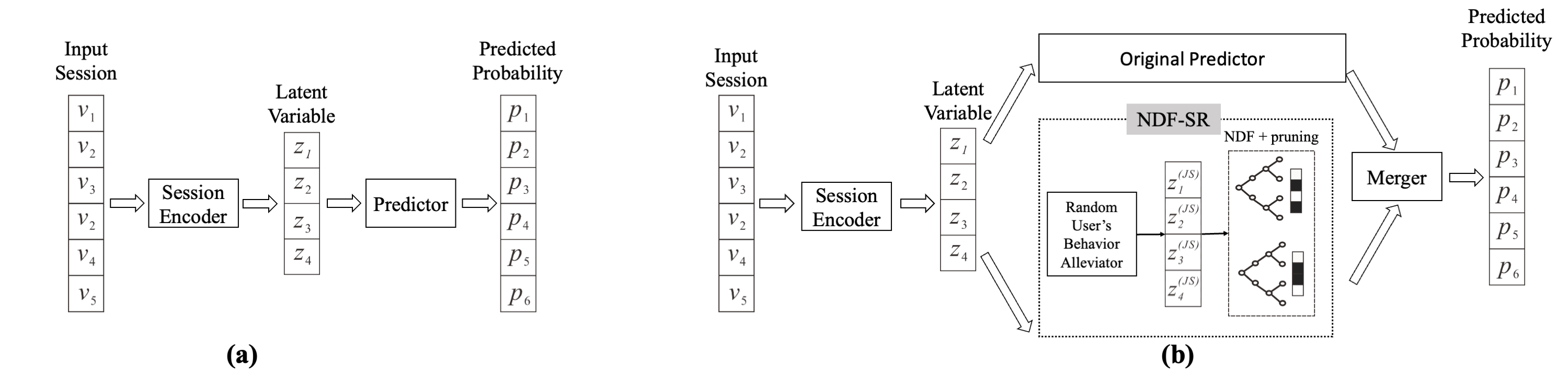}
    \caption{
    (a) The overview of the base model, (b) Framework \emph{SR-PredictAO}; Given an input session $S$, the encoder module generates the latent variable $\bm{z}$. In (a), $\bm{z}$ is passed to the base model predictor module to obtain the predicted probability distribution over all items. In (b), $\bm{z}$ is passed to both the base model predictor module and the new predictor module (called \emph{NDF-SR}) to obtain two predicted probability distributions over all items. Then, module \emph{Merger} combines the two distributions to output the final distribution.
    }
    \label{fig:model}
\end{figure*}

Due to the highly practical value in the field of modern commerce, the session-based recommendation attracts researchers' interest. In recent years, most (if not all) proposed models followed the \emph{encoder-predictor paradigm}\cite{zheng2020dgtn, chen2020handling, zhang2022dynamic, yeganegi2022star, zhang2023efficiently, pan2022collaborative}, involving 2 components. The first component is the \emph{session encoder module}, and the second component is the \emph{predictor module}. The session encoder module transforms the input session (represented in the form of a sequence of items) into an $n'$-dimensional vector called the \emph{latent variable}, where $n'$ is a positive integer denoting a model parameter. The predictor module generates a probability distribution over all items that represents how likely each item is to be the next item. The paradigm is shown in Fig. \ref{fig:model} (a).
Different existing models have different implementations 
of the encoder modules. 
For example, in \cite{SR-GNN}, the encoder module is a Gated GNN that captures complex transitions of items to obtain the latent variable, and in \cite{pan2020star}, the encoder module is a Star GNN that uses a star node,  representing the whole session, and a Highway Network, handling the overfit problem. The predictor modules of most (if not all) existing models are all \emph{linear} models.

Although existing models following the current encoder-predictor paradigm perform well, there are still some issues for further enhancement.
The first issue is that most (if not all) existing models have a \emph{low-capability} predictor module, which affects the prediction accuracy. 
Specifically, under the encoder-predictor paradigm, even though there is an advanced model in the encoder module constructing the latent variable, which could represent the latent intent of a user's purchase; it is the predictor component that makes the recommendation, which could somehow simulate the complicated decision process of a user's purchase.
Unfortunately, most (if not all) existing models use linear predictors, which are low-capability models, limiting the prediction performance.

The second issue is that designing a \emph{high-capability} model is challenging by considering the overfit problem\cite{hastie2009elements}. Specifically, one straightforward solution for the first issue is to design a high-capability model. It is well-known that an \emph{extremely} high-capability model suffers from the overfit problem. 
How to design an \emph{appropriate} high-capability model is needed for detailed investigation.

The third issue is that there is \emph{random user's behavior} in the input session, which may affect the prediction performance. It includes multi-intention problems where the user is distracted from her/his original intention of the current session. But, more generally, it can represent any random behavior of user, which could create a challenge for prediction in existing models. Previous studies \cite{pan2020star, chen2020handling} have tried to solve that in the GNN encoder but not completely.

In this paper, we propose a novel framework called \emph{\underline{S}ession-based \underline{R}ecommendation with \underline{Pred}ictor \underline{A}dd-\underline{O}n} (\emph{SR-PredictAO}). Under \emph{SR-PredictAO}, 
given an existing model called \emph{base model} in this paper, we keep all existing modules of this base model but we augment the base model with two additional modules.
The first additional module is the high-capability predictor module, which takes the latent variable as input and outputs the predicted probability distribution over all items being the next item in the session.
Maintaining the original (low-capability) predictor module, with our new high-capability predictor module, we can capture different sides of user's decision process.
The second additional module is module \emph{Merger}, which takes the probability distributions over all items predicted by both the original predictor module and the new predictor module and outputs the final probability distribution over all items.
This framework provides a lot of opportunities to researchers for optimization on how to specify these 2 modules, which is quite promising. 
The SR-PredictAO framework could be found in Fig. \ref{fig:model} (b)
where the first augmented module is named as \emph{NDF-SR} (which will be described next). 
It is worth mentioning that our framework \emph{SR-PredictAO} could be applied to all existing models following the encoder-predictor paradigm (with the two additional modules), which could further improve the prediction performance of all existing models. Due to the nature limitation that tree-based methods hardly models linear decision boundaries, we combine the tree-based model with the linear model to complement each other.

In this paper, we propose a model called \emph{\underline{N}eural \underline{D}ecision \underline{F}orest for \underline{S}ession-based \underline{R}ecommendation} (NDF-SR) for the first high-capability predictor module. Specifically, NDF-SR involves two components. 
The first component is called the \textit{random user's behavior alleviator},
which could minimize the effect of random user's behaviors for the prediction process (addressing the third issue). 
The second component is called the \emph{Neural Decision Forest} (NDF) model,
which is a high-capability model (addressing the first issue). It could be regarded as a \emph{forest} involving a number of \emph{decision trees} each constructed with the use of \emph{neural} network models.
We also propose a pruning method in the NDF model to avoid the overfit problem (addressing the second issue). 
Furthermore, in this paper, for the second \emph{Merger} module, we adopt a simple linear combination which combines
the predicted distributions from the original predictor and the new predictor to obtain the final predicted probability distribution. 
In the following, for clarify, when we describe \emph{SR-PredictAO}, we mean the framework adopting the above modules.

In summary, our contributions are shown as follows.
\begin{enumerate}
    \item To the best of our knowledge, 
    we are the first to find the important low-capability issue
    in the predictor module of most (if not all) existing models, lowering
    down their prediction accuracy. 
    
    \item  To address this important issue, we propose a framework
    called \emph{SR-PredictAO} including the high-capability 
    predictor module where this module involves two components, namely
    the \textit{random user's behavior alleviator} (addressing the random user's behavior issue) and the \emph{Neural Decision Forest} (NDF) model (addressing the low-capability predictor issue). Moreover, we propose some pruning methods in the NDF model to address the overfit problem. 
    
    \item We conduct extensive experiments on two public benchmark datasets,  namely \textit{Yoochoose} and \textit{Diginetica}, for three state-of-the-art models. Experimental results show that \emph{SR-PredictAO} improves almost all state-of-the-art models on all datasets up to 2.9\% on HR@20 (one accuracy measurement) and up to 2.1\% on MRR@20 (another accuracy measurement), which could set a new state-of-the-art in the literature. This improvement is \emph{consistent} on all datasets. By considering the consistency of improvement and the ease of applicability of our framework, we regard our contribution as a major improvement to the field of the session-based recommendation system. 
\end{enumerate}

%% file: 2-related-work.tex
\section{Related Work}\label{related}

In this section, we introduce the related work about session-based recommendation (Section \ref{SR-related}) and neural decision forest (Section \ref{NDT-related}).

\subsection{Session-based recommendation}\label{SR-related}
We categories existing studies about session-based recommendation 
into three categories: (1) conventional recommendation methods, 
(2) neural-network-based methods and (3) graph neural-network-based methods.

Due to the similarity between the \textit{session-based recommendation} (SR) problem and the traditional recommendation problem, conventional methods like Collaborative Filtering (CF) approaches~\cite{mnih2007probabilistic, koren2011advances}, nearest-neighbor approaches~\cite{davidson2010youtube, park2011session} and Markov's chain approaches~\cite{rendle2010factorizing} are applied to the SR problem.
However, due to the limited information in the session, they all performed poorly in the SR problem.

With the improvement of computation power and knowledge in \textit{Neural Network} (NN), many NN-based models, including RNN approaches \cite{hidasi2015session}, the transformer-based approach \cite{li2017neural}
and the CNN-based approach \cite{tuan20173d, yuan2019simple}, have been proposed. 
However, most of them do not perform well due to the traditional NN's encoding methods does not fit the session data well.

In recent years, graph neural networks (GNNs) have become popular and have been shown to have state-of-the-art performance in many domains. Many recommendation systems \cite{SR-GNN, chen2020handling, pan2020star, zhang2022dynamic} also utilize GNNs due to its ability of modeling complex relationships among objects. In \cite{SR-GNN}, Wu et al. apply gated graph neural networks (GGNNs) to capture the complex transitions of items, which result in accurate session representations. In \cite{chen2020handling}, to solve information loss problems in GNN-based approaches for session-based recommendation, Chen et al. proposed a lossless encoding scheme, involving a dedicatedly designed aggregation layer 
and a shortcut graph attention layer. 
In \cite{pan2020star}, Pan et al. proposed Star Graph Neural Networks with Highway Networks (SGNN-HN) for session-based recommendation. In particular, the highway networks (HN) can select embeddings from item representations adaptively to
order to prevent from overfitting. 
However, all aforementioned studies \cite{SR-GNN, chen2020handling, pan2020star, zhang2022dynamic} use the (low-capability) linear model as the predictor (described in  Section \ref{intro}).

\subsection{Tree-based method}\label{NDT-related}

The traditional tree-based method was proposed by Breiman in \cite{breiman2017classification, breiman2001random}. Its outstanding performance in simulating the human decision process is studied by Quinlan et al. in \cite{quinlan1990decision}
The high capability of the tree-based methods was shown by Mentch et al. \cite{mentch2020randomization}. With the rapid development of computation power and neural networks, a lot of effort has been made to combine classical tree-based methods with neural networks. In \cite{CNNAndRF}, Richmond et al. introduced \textit{Convolutional Neural Networks} (CNNs) as representation learners on a traditional random forest. Jancsary et al. in \cite{jancsary2012loss} introduced \emph{regression tree fields} for image restoration. To solve the problem that the traditional tree-based method cannot do backward propagation with other NN-based parts in the model, in \cite{DNDF-original}, Kontschieder et al. constructed uniform and end-to-end differentiable Deep Neural Decision Forest and applied it to some computer vision models. 
To the best of our knowledge, no existing studies about session-based recommendation
system utilizes the the tree-based models incorporated with the backward propagation with the NN-based parts in the models. We are the first one
to propose this in the field of session-based recommendation system. 

%% file: 3-preliminary.tex
\section{Preliminaries}\label{preliminaries}

In this section, we introduce 
(1) problem definition (Section~\ref{prob def}), 
(2) some preliminary knowledge about a \textit{base model}, an existing model, following the encoder-predictor paradigm (Section~\ref{base model}) and (3) the traditional version of the tree-based method (Section ~\ref{tree based method}).

\subsection{Problem Definition}\label{prob def}

The session-based recommendation is a sub-field of the next-item recommendation only with the input from a specific session. Its goal is to predict
the next item that a user will browse based on the current active session involving all previus items browsed. 
We denote $I = \{v_1, v_2, \cdots, v_{N} \}$ 
by the universal set of items in the whole dataset, where $N$ is the total number of items. A session, denoted by $\bm{s}_i = [s_{i, 1}, s_{i, 2}, \cdots, s_{i, l_i}]$, is a time-ordered sequence of items, where $i$ is a temporary index of the session, 
$l_i$ denotes the length of $\bm{s}_i$ and,
for each $t \in [1, l_i]$, 
$s_{i, t} \in I$ is the item at time step $t$ in the session. 
The goal of the session-based recommendation is to predict what the next item $s_{i, l_{i} + 1}$ is. A typical session-based recommendation system generates a probability distribution over all items predicted being the next item, i.e., $\mathbb{P}(s_{i, l_{i} + 1} | \bm{s}_i)$.

Additionally, we formally define the random-user behavior and low-capability problem that \method tries to solve as: (1) Low-capability of the predictor can be defined as low Degrees-of-Freedom (DoF) problems in the predictor because DoF usually means the max ability of the model \cite{mentch2020randomization}. (2) Random-user behavior is the mean-square difference between the real value of model encoded result and its true value \cite{stein1956variate}, the rigorous definition is in Section\ref{sec:model:predictor}. This can generally be caused by multiple comprehensive reasons including multiple intent, distractions, etc.

\subsection{Base Model}\label{base model}
\label{sess encode}

The base model (following the encoder-predictor paradigm) is formulated as follows.
\begin{align}
    &\bm{z} = f_{\textit{encode}} (\bm{s} | \Theta_{\textit{encoder}}) \\
    &\bm{y}_{base} = g_{\textit{predict}} (\bm{z} | \Theta_{\textit{predictor}})
\end{align}
where 
(1) $\bm{s}$ is the input session (represented in the form of a sequence of items), 
(2) $\bm{z}$ is the latent variable generated by the encoder module of the model,
(3) $\bm{y}_{base}$ denotes the probability distribution over all items predicted being the next item, 
(4) $f_{\textit{encode}}$ is the encoder which takes the input session as input and outputs a latent variable (a vector in $\mathbb{R}^{n'}$)
(5) $g_{\textit{predict}}$ is the predictor module which takes the latent variable as input and outputs the probability distribution, and 
(6) $\Theta_{\textit{encode}}$ ($\Theta_{\textit{predict}}$) is the parameter configuration of the encoder (predictor) module.

As described in Section~\ref{intro}, different existing models have different implementations 
of the encoder modules. 
In the following, we describe the encoder module and the predictor module
of a base model of some state-of-the-art models.

\subsubsection{Encoder Module}

This section focuses on the most popular base model's session encoding method, the GNNs encoder. But our methods can work on all kinds of session encoders as long as it generates a latent variable.
GNNs are \textit{Neural Networks} (NN) that directly operate on the graph of data, given a graph $G = (V, E)$, where each node $v_i \in V$ represents an item in $\bm{s}$ (the session). Typically, $v_i$ is associated with a node feature vector $\bm{x}_i$, which is the input to the first layer of GNNs. $\bm{x}_i \in \mathbb{R}^n $ is obtained by multiplying the embedding matrix (we define embedding matrix as $\bm{A} \in \mathbb{R}^{N \times n} $ with the item ID), where $n$ is the embedding dimensionality. And $\bm{A}$ is a trainable matrix. Assume we totally have $L$ layers of GNN. The formula of $l$-th ($l \leqslant L$) layer of GNN can be represented as follows:
\begin{align}
    & \bm{x}_i^{(l+1)} = f^{(l)} (\bm{x}_i^{(l)}, \bm{a}_i^{(l)}) \\
    & \bm{a}_i^{(l)} = agg^{(l)} ( \{msg^{(l)} (\bm{x}_i^{(l)}, \bm{x}_j^{(l)}) | (j, i) \in E_{in}(i) \})
\end{align}
where $\bm{x}_i^{(l)}$ is the embedding vector of node $i$ in the $l$-th layer of the GNN, and $E_{in}(i)$ is the set of incoming edges for node $v_i \in V$. The message processing function at the $l$-th layer $f^{(l)}$ generates the updated embedding of the target node based on its neighborhood. $agg^{(l)}$ is the aggregate function that connects the information of different edges together, and $msg^{(l)}$ is the message-extracting function that obtains information from the edge between $(x_i^{(l)}, x_j^{(l)})$. Let $L$ be the total number of layers in the GNN. After $L$ steps of message passing, the final representation for the latent variable is:
\begin{equation}
    \bm{h}_G = f_{out} (\{\bm{x}_i^{(L)} |v_i \in V \} )
\end{equation}
$\bm{h}_G$ is the graph-level representation that we regard as the graph latent variable generated by the readout function $f_{out}$. 

After the graph level latent variable $\bm{h}_G$ is obtained, most models adds some additional information to obtain a better result. For example, \cite{chen2020handling} adds all results of the Embedding layer, EOPA Layer, and SGAT Layer's (two special kinds of GNN mentioned in \cite{chen2020handling}) information to the graph representation, and \cite{pan2020star} formulates the final result by concatenating $\bm{z}_g$ and $\bm{z}_r$, which are the last item's representation and the combination of all the graphs' result representation come from different levels respectively. After considering all the required information of the base model, we define this vector as the latent variable $\bm{z} \in \mathbb{R}^{n'} $, where $n'$ is the dimensionality of the latent variable. This approach is used in almost all well-known session-recommendation models \cite{pan2020star,zhang2022dynamic,SR-GNN,chen2020handling} .

\subsubsection{Predictor Module}
After the encoder module outputs the latent variable, 
the predictor module takes this as input and performs the 
following steps. 
\begin{enumerate}
    \item The first step is to perform a prediction function (normally a linear model), which takes the latent variable as input and outputs an embedding called the \emph{session embedding} $\bm{s}_h \in \mathbb{R}^n$
    where $n$ is the dimensionality of the session embedding, same as the embedding dimension of $\bm{A}$
        \begin{equation}
            \bm{s}_h = \text{Linear}(\bm{z}) 
        \end{equation}
    \item The second step is to obtain the \emph{score vector} $\bm{c} \in \mathbb{R}^{N}$ over all items
    predicted being the next item. 
        \begin{equation}
            \bm{c} = [c_1, c_2, \cdots, c_{N}]^T = \bm{A} \bm{s}_h
        \end{equation}        
        where $c_i \in \mathbb{R}$ is a score of item $v_i$ predicted being the next item for each $i \in [1, N]$
        and $\bm{A} \in \mathbb{R}^{N \times n}$ is the item embedding matrix we used before.
    \item The third step is to obtain the \emph{probability vector} $\hat{\bm{y}}_{base} \in \mathbb{R}^{N}$
    over all items predicted being the next item by using the softmax function based on the score vector $\bm{c}$. 
        \begin{equation}
            \hat{\bm{y}}_{base} = softmax(\bm{c})  = \frac{\exp(\bm{c})}{\sum_{i \in [1, N]} \exp(c_i)} 
        \end{equation}

\end{enumerate}

\subsection{Tree-based method}\label{tree based method}

From the mathematical point of view, the tree-based method is a way of generating a locally constant function, represented by function $tree: \mathbb{R}^{n'} \rightarrow \mathbb{R}^N$ that divides the input space $\mathbb{R}^{n'}$ into many regions, and give each subspace a constant value in $\mathbb{R}^N$. And we can define the tree recursively by first defining the \textit{tree-split} function $\varphi$:
\begin{equation}\label{simpFunc}
    \varphi(\bm{x}) = \chi(\bm{x} \in S) \bm{c}_l+ \chi(\bm{x} \notin S) \bm{c}_r
\end{equation}
where $S \subseteq \mathbb{R}^{n'}$ is a subregion of the input space, and $\chi(\bm{x} \in S)$ is judging function that returns 1 when $x \in S$, and 0 otherwise. The $\bm{c}_l, \bm{c}_r$ are defined as the left and right nodes of the tree-split. 
If $\bm{c}_l$ or $\bm{c}_r$ have its value in $\mathbb{R}^N$, where $N$ is the dimension of the predicted result, then we say it is a \textit{leaf node}; if not, it is an \textit{internal node} that is associated with another tree split $\varphi_{l/r}$. 
And the tree function can be represented as $tree(\bm{x}) = \varphi_{root} (\bm{x})$ where $\varphi_{root}$ is the tree-split function associated with the \textit{root node}, the beginning node of the tree. The max number of tree-split need to have from the root to the leaf node is defined as \textit{depth}.

For example, in Fig. \ref{fig:NDT}, each node $d_i (i \in [1, 7])$ is associated with a tree split function $\varphi_i$ with corresponding region $S_i$. The node of $d_1$ is the root node (i,e., $tree = \varphi_{1}$), the nodes of $d_{i \neq 1}$ are internal nodes. And node of $\bm{\pi}_j (j \in [1, 8])$ is leaf node that have its value $\bm{\pi}_j \in \mathbb{R}^{N}$.

%% file: 4-model.tex
\section{Framework SR-PredictAO}\label{ourEnhancement}

Framework \emph{SR-PredictAO} involves two modules, namely the high-capability
predictor module (Section~\ref{sec:model:predictor}) and the Merger module (Section~\ref{sec:model:merger}). 
The  training process of \emph{SR-PredictAO} is presented in Section~\ref{sec:training}. 

\subsection{High-Capability Predictor Module}
\label{sec:model:predictor}

We propose a model called \emph{\underline{N}eural \underline{D}ecision \underline{F}orest for \underline{S}ession-based \underline{R}ecommendation} (NDF-SR) for the high-capability predictor module. Specifically, NDF-SR involves two components. 
The first component is called the \textit{random user's behavior alleviator} (Section~\ref{JS-Estimator}) and the second component is called the \emph{Neural Decision Forest} (NDF) model (Section~\ref{sec:NDF}).
As described in Section~\ref{intro}, we also propose a pruning method in the NDF model to avoid the overfit problem. This pruning method could be found in the description for the second component.

\subsubsection{Random User's Behavior Alleviator}\label{JS-Estimator}

The base-model encoded latent variable for the previous session view of items is normally heavily affected by random user's behavior. To solve this problem, we could take the Empirical Bayes' point of view \cite{stein1956variate}. For Empirical Bayes', the observed data is not the underlying true value but a sample under a certain distribution around the truth. We would design our Alleviator under this cognition.

Formally, if a batch $\bm{Z} \in \mathbb{R}^{m \times n'}$ of $m$ latent variables each with dimensionality of $n'$ we observe from the base model's encoder is:
\[\bm{Z} = 
\begin{bmatrix}
    \bm{z}_1^T \\
    \bm{z}_2^T \\
    \vdots \\
    \bm{z}_m^T
\end{bmatrix} = 
\begin{bmatrix}
    \bm{\xi}_1 \\
    \bm{\xi}_2 \\
    \vdots \\
    \bm{\xi}_{n'}
\end{bmatrix}^T
=
\begin{bmatrix}
  z_{11} & z_{12} & \cdots & z_{1n'} \\
  z_{21} & z_{22} & \cdots & z_{2n'} \\
  \vdots & \vdots & \ddots & \vdots \\
  z_{m1} & z_{m2} & \cdots & z_{mn'}
\end{bmatrix} 
\]
We denote $\bm{z}_j$ to be the $j$-th row of $\bm{Z}$ and also the latent variable of the $j$-th session in the batch for each $j \in [1, m]$. We denote $\bm{\xi}_i$ to be the $i$-th column of $\bm{Z}$ for each $i \in [1, n']$.

$\bm{Z}$ is not the underlying truth value for the latent variable but a sample from a distribution with the underlying truth value as its expected value. Suppose that 
$\bm{\mu} \in \mathbb{R}^{m \times n'}$ denotes the correspondence truth values as follows.
\[\bm{\mu} = 
\begin{bmatrix}
    \mu_{11} & \mu_{12} & \cdots & \mu_{1n'} \\
    \mu_{21} & \mu_{22} & \cdots & \mu_{2n'} \\
    \vdots & \vdots & \ddots & \vdots \\
    \mu_{m1} & \mu_{m2} & \cdots & \mu_{mn'}
\end{bmatrix} = 
\begin{bmatrix}
    \bm{\mu}_1^T \\
    \bm{\mu}_2^T \\
    \vdots \\
    \bm{\mu}_m^T
\end{bmatrix} 
\]

The Empirical Bayes' assumption is that $\forall i, j; z_{ij} | \mu_{ij} \sim \mathcal{N}(\mu_{ij}, \sigma_j^2)$, which is a normal distribution with mean $\mu_{ij}$ and variance $\sigma_j^2$, with an additional assumption that $\sigma_j^2 \geqslant 1$. This assumption also means that the variance is the same across different columns. We aim to obtain an estimator for $\bm{\mu}$ given the observation $\bm{Z}$. The \textit{\underline{M}aximum \underline{L}ikelihood \underline{E}stimator} (MLE) that is commonly used in the field suggests that we should just take the $\bm{Z}$ itself. That is, for each $i \in [1, m]$ and each $j = [1, n']$,
\begin{equation}
    \hat{\mu}_{ij}^{(MLE)} = z_{ij} 
\end{equation}

But, our alleviator uses the \textit{\underline{J}ames-\underline{S}tein \underline{E}stimator for \underline{S}ession-based \underline{R}ecommendation} (JSE-SR) that applies indirect evidence from other values of the same entry in the batch. The JSE-SR is defined as follows: 

\begin{equation}\label{eq:JSResult}
    \hat{\mu}_{ij}^{(JS)} = (1 - \frac{m - 2}{\|\bm{\xi}_j\|^2}) z_{ij}
\end{equation}

For each of the two estimators $\hat{\mu}_{ij}$ (i.e.,$\hat{\mu}_{ij}^{(MLE)}$ and $\hat{\mu}_{ij}^{(JS)}$), the effect of random user's behavior on the latent variable can be quantified as follows. For each $j \in [1, n']$, $\mathbb{E} [\sum_{i = 1}^m (\mu_{ij} - \hat{\mu}_{ij})^2]$. 

We can show the following lemma. In this lemma, we know that the estimator $\hat{\mu}_{ij}^{(JS)}$ gives a smaller error compared with the estimator $\hat{\mu}_{ij}^{(MLE)}$.
\begin{lemma}
\begin{equation}
\mathbb{E}[\sum_{i = 1}^m (\mu_{ij} - \hat{\mu}^{(JS)}_{ij})^2] \leqslant \mathbb{E}[\sum_{i = 1}^m (\mu_{ij} - \hat{\mu}^{(MLE)}_{ij})^2]    
\end{equation}
\end{lemma}
\emph{Proof Sketch:}
Firstly, for all predictor $\hat{\mu}_{ij} := \hat{\mu}_{ij}(z_{ij})$ of $\mu_{ij}$, we can decompose $\mathbb{E}[\sum_{i = 1}^m (\mu_{ij} - \hat{\mu_{ij}})^2] = \sum_{i = 1}^m \mathbb{E}[(z_{ij} - \hat{\mu}_{ij})^2] + 2\sum_{i = 1}^m \mathbb{E} [(\hat{\mu}_{ij} - \mu_{ij})(z_{ij} - \mu_{ij})]$.
Secondly, we perform integration by parts, we have: $\mathbb{E}[(z_{ij} - \mu_{ij})(\hat{\mu}_{ij} - \mu_{ij})] = \sigma_j^2 \mathbb{E}[\frac{\partial \hat{\mu}_{ij}}{\partial z_{ij}}]$.
Thirdly, we plug the $\hat{\mu}_{ij}^{(JS)}$ and $\hat{\mu}_{ij}^{(MLE)}$ into the equation, we have Equation~\ref{eq:JSResult}. A complete proof can be found in Appendix~\ref{JS-prove}
 \done

Therefore, applying JSE-SR to all entries in $\bm{Z}$, we have: 
\begin{equation}
\hat{\bm{Z}}^{(JS)} = [\hat{\mu}^{(JS)}_{ij}]_{i \in [1, m], j \in [1, n']}
\end{equation}

\subsubsection{Neural Decision Forest (NDF)} 
\label{sec:NDF}

As described in Section~\ref{intro}, 
the Neural Decision Forest (NDF) model 
could be regarded as a \emph{forest} involving a number of \emph{decision trees} each constructed with the use of \emph{Neural Network} (NN) models. Each decision tree in this model
is formally named as a \emph{Neural Decision Tree} (NDT). 

In the following, we first define NDT and then NDF.

\medskip
\noindent
\textbf{NDT:}
\label{NDT}
Our proposed NDT method is the part that provides (more than) enough capability to solve the lack of capability problem of the linear predictor.
Considering the representation learning in the session-based recommendation, our proposed NDT  differs from the traditional trees that greedily find the split that may reduce the loss function in the given variable space and entries proposed by \cite{breiman2017classification}, which requires a fixed encoder, but our proposed NDT uses NN to do the split and are optimized by backward propagation together with the encoder. In our case, this encoder is normally a GNN-based encoder. The NDT that has depth $d$, and it takes values from alleviator-processed latent variable $\bm{z}^{(JS)} \in \mathbb{R}^{n'}$ as input. It consists of the following. 

\begin{itemize}
    \item A decision function (normally a deep neural network): $f: \mathbb{R}^{n'} \rightarrow \mathbb{R}^{2^d-1}$ (because a tree with depth $d$ requires $2^d - 1$ number of the split, resulting in $2^d$ leaf nodes)
    \item A probability score matrix $\bm{\pi} \in \mathbb{R}^{2^d \times N}$ (which is trainable) for all leaf nodes: 
    \begin{equation}
        \bm{\pi} = [\pi_{ij}] = [\bm{\pi}_1, \cdots, \bm{\pi}_{2^d}]^T
    \end{equation}
    We mark the leaf nodes of a tree from left to right with index $1, 2, \cdots, 2^d$, where the $i$-th leaf node means the leaf node with index $i$. Note that under our definition, the NDT is always  a balanced tree. $\pi_{ij}$ means the probability score of the $j$-th item in the $i$-th leaf node.  $\bm{\pi}_i$ means a vector containing the probability scores of all items in $I$ of the $i$-th leaf node.
\end{itemize}

\begin{figure}[hpbt]
    \centering
    \includegraphics[width=0.8\linewidth]{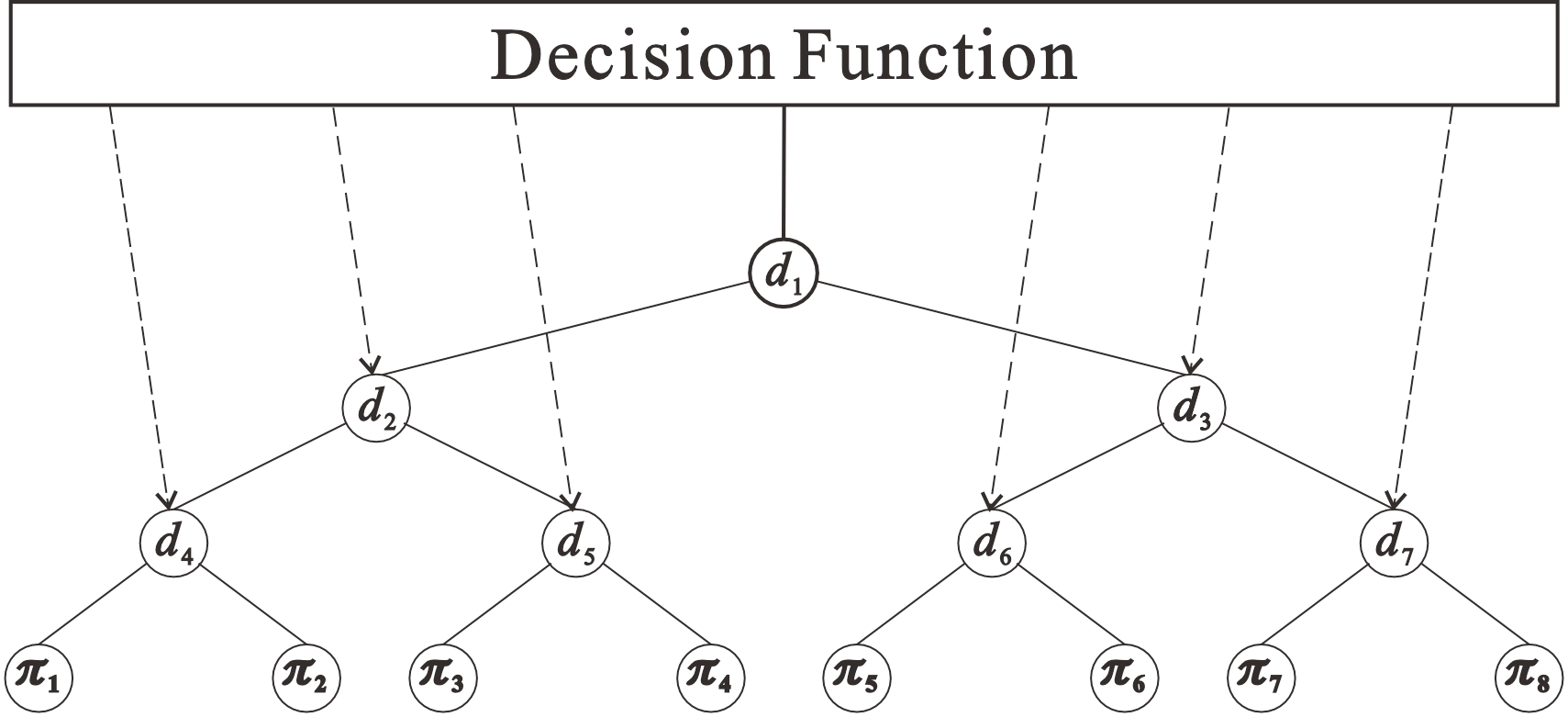}
    \caption{The overview of the NDT, decision function gives the split score for root and internal nodes, and the leaves nodes' result is the probability of the session reaching the node}
    \label{fig:NDT}
\end{figure}

The NDT works as follows. The decision function generates a decision score for each split. Then, applying a sigmoid function to the decision score to obtain the right and left decision probability. A binary split is associated with the probability of arriving at the root of this split as $p_{root}$, which is generated by previous splits. Let $s = \sigma(f(z^{(JS)})) $. The split here means the process of giving an item in the root of the subtree what is the probability that this item goes to the right and left of the root. The probability is calculated as follows.

\begin{equation}
    \begin{cases}
        p_{left} = p_{root} \cdot s \\
        p_{right} = p_{root} \cdot (1-s) 
    \end{cases}
\end{equation}
For example, in Fig. \ref{fig:NDT}, $p_{root}$ for node $d_1$ is 1, and $p_{root}$ for node $d_2$ is set to $p_{left}$ computed within node $d_1$.

We recursively apply this split method from the tree's root to the leaf nodes. We obtain the \textit{}{leaf-reaching probability} $\bm{p}_{leaf} = [p_1^{(leaf)}, p_2^{(leaf)}, \cdots, p_{2^d}^{(leaf)}]^T \in \mathbb{R}^{2^d}$ to represent what is the probability that this session may fall into each leaf node. Then, multiply $softmax(\bm{\pi})$ matrix by $\bm{p}_{leaf} $ to obtain the probability distribution $\hat{\bm{p}} \in \mathbb{R}^{N}$ over all items that this session may represent.
\begin{equation}
    \hat{\bm{p}} = \bm{p}_{leaf}^T softmax(\bm{\pi}) = \sum_{k = 1}^{2^d} p_k^{(leaf)} softmax(\bm{\pi}_k) 
\end{equation}
where $\hat{\bm{p}}$ is the predicted probability for each item for this tree. To make $\bm{\pi}$ normalized, we apply the softmax function before we use it.

\medskip
\noindent
\textbf{Pruning:}
\label{NDT Pruning}
Because that all tree-based methods, including NDT, suffer from serious overfitting because they normally have excessive capability. The problem is more severe in our case since our NDT is trained simultaneously with the encoder. To solve that problem, we propose \textit{NDT-pruning} that can control the excessive capability to control overfitting.

Traditional pruning uses the judgment of loss function to see which leaves should drop, but for an NDT, it is hard to do a similar thing. Thus, to prune the NDT, we apply a random mask to the outcomes of NDT. So we do the following:
\begin{equation}
    \bm{p}'_{leaf} = softmax(RandomMask(\bm{p}_{leaf}, r))
\end{equation}
where $\bm{p}_{leaf} \in \mathbb{R}^{2^d} $ is the leaf-reaching probability, and each leaf node has a probability $r$ (we call it pruning rate) to be 0, and $r \in [0, 1]$. After the random mask, we use $\bm{p}'_{leaf}$ to replace $\bm{p}_{leaf}$ 
to obtain $\hat{\bm{p}}$, the predicted next-item distribution of this tree.

Since NDT typically has excessive capability than needed, which may fit into unrelated information in data, this makes the model easy to overfit. Our proposed NDT-pruning controls the overfitting by removing the excessive capability of the NDT. By choosing a good pruning rate, we can control the capability of our model in a reasonable range that can compensate for the lack of capability in linear predictors and not be too high to overfit. More details of the relation between the model's capability and NDT-pruning can be found in Section~\ref{sec:dof}

\medskip
\noindent
\textbf{NDF:}
\label{NDF}
We construct the NDF by the basic building block NDT and NDT-pruning in this section. Breiman proved that combining trees into a forest model generally makes the model's outcome more stable \cite{breiman2001random}. Non-neural trees that formulate Random Forest should have a different mask of entries for every split, but that is not possible if we use a uniform decision function for each tree. So, we independently drop some entries for each NDT.

For example, if an input Alleviator-processed latent variable for the NDT is 
$\bm{z}^{(JS)} = [z_1^{(JS)} \cdots, z_{n'}^{(JS)}]^T \in \mathbb{R}^{n'}$
, for the $i$-th NDT after the variable mask-off, a fixed subset of $\bm{z'}_i = [z_{1_i}, z_{2_i}, \cdots, z_{\gamma_i}]$ where $|\bm{z}_i'| = \gamma_i \leqslant n'$, and $\bm{z'} \subseteq \bm{z}^{(JS)}$. For each NDT, the list of entries to drop is randomly selected when building the model, but this list is fixed during training. If there are $T$ number of NDTs in the NDF-SR, and their predicted next-item probability is $\bm{P} = [\hat{\bm{p}}_1, \hat{\bm{p}}_2, \cdots, \hat{\bm{p}}_T]$, where $\hat{\bm{p}}_i \in \mathbb{R}^N$ for all $i = 1, 2, \cdots, T$. The NDF's predicted result is:
\begin{equation}
    \hat{\bm{y}}_{NDF-SR} = \frac{1}{T} \cdot (\sum_{i = 1}^T \hat{\bm{p}}_i)
\end{equation}
which is also the predicted result of the NDF-SR, our proposed high-capability predictor.

\medskip
\noindent
\textbf{Time complexity analysis:} Under tree-parallel setting, the NDF-SR module's time complexity is $\mathcal{O} (m \cdot n' \cdot k + m \cdot k \cdot N)$, where $k$ is the number of leaves in the tree (typically 32 to 64). This is only slightly higher than the $\mathcal{O}(m \cdot n' \cdot N)$ for traditional linear predictors.

\subsection{Merger Module}
\label{sec:model:merger}

In this paper, for the second \emph{Merger} module, we adopt a simple linear combination which combines
the predicted distributions from the original predictor and the new predictor to obtain the final predicted probability distribution by using a user parameter $q \in [0, 1]$ as follows. 
\begin{equation}
    \hat{\bm{y}} = q \cdot \hat{\bm{y}}_{base} + (1 - q) \cdot \hat{\bm{y}}_{NDF-SR}
\end{equation}
Here, $\hat{\bm{y}} \in \mathbb{R}^{N}$ is the probability distribution over all items predicted being the next item (which is the combined result from the original predictor module
and the new predictor module). $\hat{\bm{y}}$ is the output of framework \emph{SR-PredAO}.

\subsection{Training}
\label{sec:training}

Note that $\hat{\bm{y}}$ obtained in module Merger is the output of framework \emph{SR-PredAO}. Let $\bm{y}$ be the real probability distribution over all items being the next item, which is a one-hot vector. The loss function of framework \emph{SR-PredAO} $\mathcal{L}(\cdot, \cdot)$ is the same as the one used in the base model, which is the cross-entropy loss.
\begin{equation}
    \mathcal{L}(\bm{y}, \hat{\bm{y}}) = -\bm{y}^T\log(\hat{\bm{y}})
\end{equation}
 For initialization, all trainable parameters in both the base model and the
additional modules in framework \emph{SR-PredAO} are initialized randomly, and they are jointly
updated in an end-to-end back propagation manner.

%% file: 5-experiments.tex
\section{Experiment}\label{experiment}

We give the experiment setup in Section~\ref{subsec:exp:setup} and the results in Section~\ref{subsec:exp:result}. Implementation of this paper can be found in \href{our GitHub repo}{https://github.com/RickySkywalker/SR-PredictAO-official}

\subsection{Experimental Setup}
\label{subsec:exp:setup}

\subsubsection{Datasets}
We evaluated the performance of state-of-the-art models and our proposed framework on the following two benchmark real-world datasets:
\begin{itemize}
    \item \textit{Yoochoose}\footnote{http://2015.recsyschallenge.com/challenge.html} is a dataset obtained from the RecSys Challange 2015, which contains user sessions of click events from an online retailer.
    \item \textit{Diginetica}\footnote{http://cikm2016.cs.iupui.edu/cikm-cup} is a dataset released by the CIKM Cup 2016, which includes user sessions extracted from e-commerce search engine logs.
\end{itemize}

Our dataset preprocess directly following \cite{pan2020star, chen2020handling, li2017neural}. The statistics of the datasets after pre-processing are provided in Table \ref{tab:stats}.

\begin{table}[h]
    \centering
    \begin{tabular}{lll}
        \toprule
        \textbf{Statistic} & \textbf{Yoochoose 1/64} & \textbf{Diginetica}  \\
        \midrule
        \# of Clicks & \num{565332} & \num{982961}  \\ 
        \midrule
        \# of Training Sessions & \num{375625} & \num{647523} \\ 
        \midrule
        \# of Test Sessions & \num{55896} & \num{71947} \\ 
        \midrule
        \# of Items & \num{17792} & \num{43097}\\ 
        \midrule
        Average length & 6.14  & 5.12 \\
        \bottomrule
    \end{tabular}
    \caption{Statistics of datasets}
    \label{tab:stats}
\end{table}

\subsubsection{Evaluation Metrics}

Following previous studies \cite{NARM, RepeatNet, SR-GNN, FGNN, chen2020handling, pan2020star, zheng2020dgtn, zhang2022dynamic}, we adopt the commonly used HR@20 (Hit Rate)\footnote{Note that \cite{NARM, RepeatNet, SR-GNN, FGNN, chen2020handling, pan2020star, zheng2020dgtn, zhang2022dynamic} used different metric names for HR@20 (e,g, P@20 and Recall@20). But, they used the same formula to obtain this measurement (i.e., the proportion of cases when the target item is in the top-20 items in all test cases).} and MRR@20 (Mean Reciprocal Rank) as our evaluation metrics.

\subsubsection{Base Model \& Baselines}

Framework \emph{SR-PredAO} involves a base model (together with our proposed high-capability predictor module and the Merger module). In our experiments, we choose the following three base models, namely LESSR~\cite{chen2020handling}, SGNN-HN~\cite{pan2020star} and DIDN~\cite{zhang2022dynamic}, since they are representative in the literature. Roughly speaking, LESSR has a clear encoder-predictor paradigm for the ease of illustration and conducting subsequent experiments. SGNN-HN and DIDN have the best performance on datasets Yoochoose 1/64 and Diginetica, respectively.

We also considered using some newer proposed models like \cite{yeganegi2022star, zhang2023efficiently, pan2022collaborative} as base models, but their performance is less satisfactory in our benchmarks and thus not demonstrated in the paper. In the following, when we describe framework \emph{SR-PredAO} using the base model $M$, we write \emph{SR-PredAO(M)}. 

\begin{figure*}[t!]
    \centering
    \includegraphics[width=0.9\textwidth]{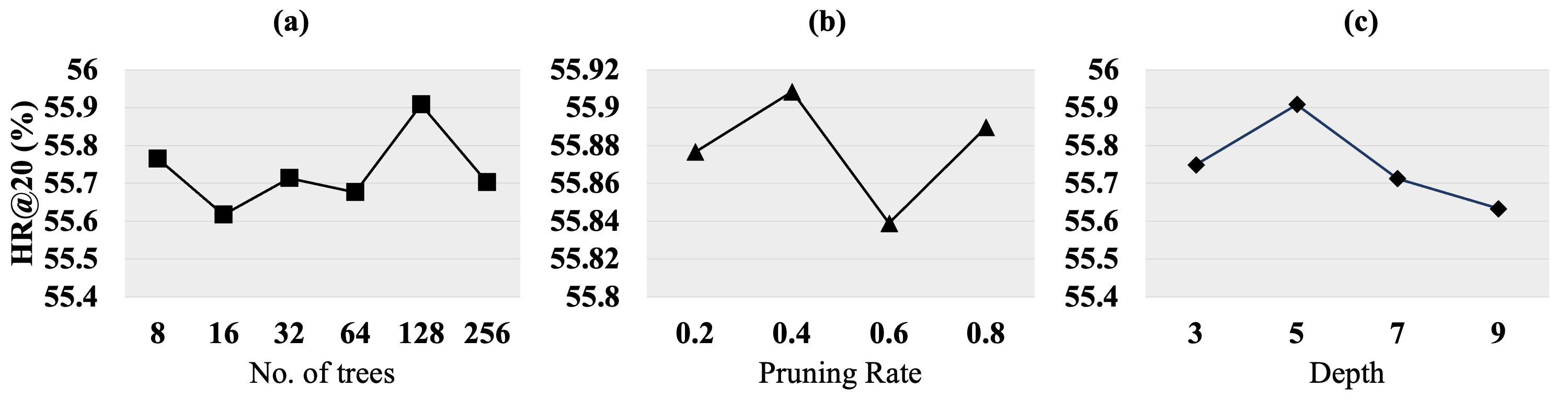}
    \caption{The hyper-parameter study results of SR-PredAO(SGNN-HN)}
    \label{fig:hyper}
\end{figure*}
\begin{figure*}[t!]
    \centering
    \includegraphics[width=0.9\textwidth]{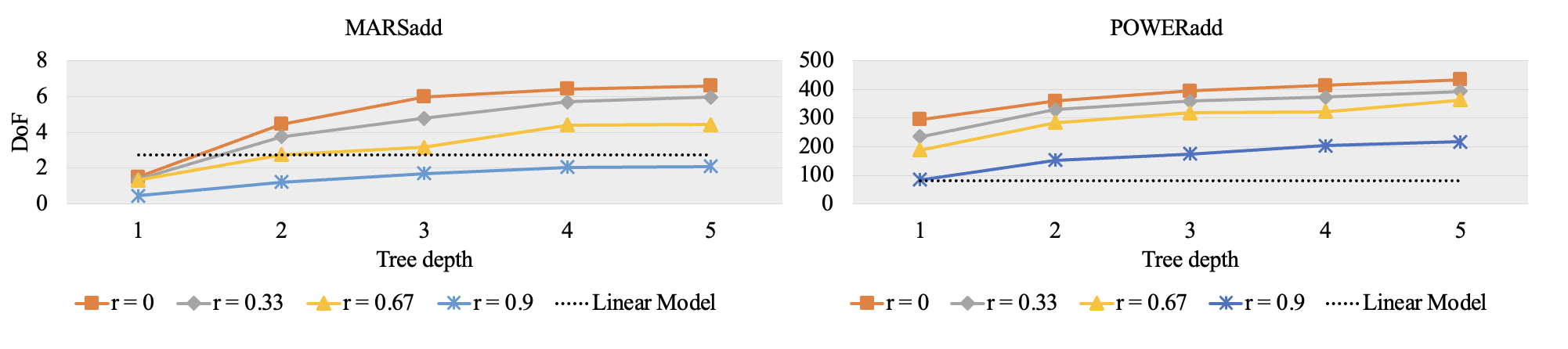}
    \caption{DoF of NDF-SR with different depth and pruning rates (r), dotted lines represent DoF of linear model}
    \label{fig:dof}
\end{figure*}

To more effectively illustrate the superiority of our framework's capability in enhancing models. In addition to three base models without enhancement as baselines, we have also included other baselines in the comparison. These encompass traditional recommendation methods such as \textbf{Item-KNN} \cite{sarwar2001item}, the GRU-based method \textbf{GRU4REC} \cite{hidasi2015session}, two transformer-like methods, \textbf{STAMP} \cite{liu2018stamp} and \textbf{SR-IEM} \cite{pan2020rethinking}, and a basic GNN-based method, \textbf{SR-GNN} \cite{SR-GNN}.

\subsubsection{Implementation Details}
In the \emph{SR-PredAO} framework, hyper-parameters (e.g., batch size and learning rate) for the base models are kept as the best experimental configurations reported in their papers\cite{zhang2022dynamic, pan2020star, chen2020handling}. This allows us to observe the improvements made by the \emph{SR-PredAO} framework, which includes the new predictor module and the merger module, over the base models. The additional hyper-parameters are determined through binary search. Additionally, every reported result is the best outcome for both the baselines and the \method-enhanced models, and these configurations may vary. We also use the accuracy of the training data as the validation set for model selection. The training averagely costs 3 RTX-4090 GPU days.

\subsection{Experimental Results}
\label{subsec:exp:result}

\subsubsection{Performance Comparison}

\begin{table}[h]
    \centering
    \resizebox{\columnwidth}{!}{\begin{tabular}{ccccc}
     \toprule
     \multirow{2}{*}{\textbf{Method}} & \multicolumn{2}{c}{\textbf{Diginetica}} & \multicolumn{2}{c}{\textbf{Yoochoose 1/64}} \\
     \cline{2-5}
                            &{\bf HR@20}    & {\bf MRR@20}  & {\bf HR@20}   & {\bf MRR@20}\\
     \midrule
     Item-KNN               & 35.75         & 11.57         & 51.60         & 21.81 \\
     GRU4REC                & 29.45         & 8.33          & 60.64         & 22.89 \\
     STAMP                  & 45.64         & 14.32         & 68.74         & 29.67 \\
     SR-IEM                 & 52.35         & 17.64         & 71.15         & 31.71 \\
     SR-GNN                 & 50.73         & 17.59         & 70.57         & 30.94 \\
     \midrule
     LESSR                  & 51.71             & 18.15             & 70.94                     & 31.16 \\
     SR-PredAO(LESSR)       & \textbf{53.10}    & \textbf{18.38}    & \textbf{71.73}            & \textbf{31.70} \\
     \cline{2-5}
     Improvement (\%)       & \textbf{2.7}      & \textbf{1.3}      & \textbf{1.1}              & \textbf{1.7} \\
     p-value                & $\bm{<10^{-5}}$   & -                 & $\bm{1.8\times10^{-3}}$   & -   \\
     \midrule
     SGNN-HN                & 55.67             & \textbf{19.12}    & 72.06                     & \textbf{32.61} \\
     SR-PredAO(SGNN-HN)     & \textbf{55.91}    & 19.06             & \textbf{72.62}            & 32.47 \\
     \cline{2-5}
     Improvement (\%)       & \textbf{0.4}      & -0.3              & \textbf{0.8}              & -0.4 \\
     p-value                & $0.179$           & -                 & $\bm{1.8\times10^{-2}}$   & -   \\
     \midrule
     DIDN                   & 56.22             & 20.03             & 68.95                     & 31.27 \\
     SR-PredAO(DIDN)        & \textbf{57.86}    & \textbf{20.49}    & \textbf{69.50}            & \textbf{31.44}  \\
     \cline{2-5}
     Improvement (\%)       & \textbf{2.9}      & \textbf{2.3}      & \textbf{0.8}              & \textbf{0.5} \\
     p-value                & $\bm{<10^{-5}}$   & -                 & $\bm{2.3\times10^{-2}}$   & -   \\
     \bottomrule
    \end{tabular}}
    \caption{Experimental result (\%) on three enhanced models and baselines on two datasets}
    \label{tab:results}
\end{table}

Table~\ref{tab:results} shows the experimental results for all models, we can see that \method has a relatively significant improvement on HR@20 for all models and on MRR@20 for almost all models. Specifically, framework \emph{SR-PredAO}, when applied to existing state-of-the-art models,could have up to 2.9\% improvement on HR@20 and 2.3\% of improvement on MRR@20. Furthermore, to test the significance of our \method, we conducted the two-proportion z-tests and reported p-value of such test in the table. We can spot that when the base model has a good encoder-predictor split (i.e., DIDN and LESSR). The \method enhancement statistically significantly outperforms the base models. According to \href{https://paperswithcode.com/}{\textcolor{blue}{Paper-with-code}}, \textit{SR-PredAO} 
 achieves state-of-the-art on all experimented benchmarks on HR@20 and almost all on MRR@20.

It is worth mentioning that using  framework \emph{SR-PredAO} 
on any existing model could automatically improve the prediction accuracy, which is a great advantage. Compared with recent papers \cite{chen2020handling, pan2020star, zhang2022dynamic, yeganegi2022star, zhang2023efficiently, pan2022collaborative, zheng2020dgtn} showing that 1.4\% of improvement is considered as a major contribution, framework \emph{SR-PredAO} has a significant improvement in the field.

\subsubsection{Ablation Studies}\label{Abl}
This section presents the ablation studies results for two important components in \method, namely Random User Behavior Alliviator and NDT-pruning

\begin{table}[h]
    \centering
    \begin{tabular}{ccccc}
        \toprule

        \textbf{Dataset}                     & \textbf{type}    & \textbf{LESSR}    & \textbf{SGNNHN}   & \textbf{DIDN} \\
        \midrule
        \multirow{3}{*}{\textbf{Diginetica}} & full model       & \textbf{53.15}    & \textbf{55.91}    & \textbf{57.86} \\
                                             & w/o Alliviator   & 52.99             & 55.79             & 57.33 \\
                                             & w/o Pruning      & 53.06             & 55.78             & 57.26 \\
        \midrule
        \multirow{3}{*}{\textbf{YC 1/64}}    & full model       & \textbf{71.73}    & \textbf{72.62}    & \textbf{69.50} \\
                                             & w/o Alliviator   & 71.67             & 72.58             & 69.26 \\
                                             & w/o Pruning      & 71.66             & 72.58             & 69.20 \\

        \bottomrule
    \end{tabular}
    \caption{Ablation test results (\%) on random user's behavior alleviator (Alleviator) and NDT-pruning (Pruning)}
    \label{tab:eff_JS-Pru}
\end{table}

Table~\ref{tab:eff_JS-Pru} shows that if we drop the random user's behavior alleviator or NDT-Pruning in framework \emph{SR-PredAO}, the improvement of \emph{SR-PredAO} over the base model drops to a great extent in the Diginetica dataset but not that much in \textit{YooChoose} (YC) 1/64 dataset. 
This is because the YC dataset is simpler compared with Diginetica. And in YC, the random user behavior and the overfitting problem are not that obvious to a certain extent.

\subsubsection{Hyper-parameter Study}

In this section, we study how the number of trees, the depth of the tree, and the pruning rate affect the performance of \emph{SR-PredAO}. All the results are shown in Fig. \ref{fig:hyper}. When the number of trees reaches 128, HR@20 of \emph{SR-PredAO} is the highest. When the number of trees is larger 128, HR@20 decreases because more trees affects the model's learning capacity. For the pruning rate, as long as we do not remove the pruning feature, we can see that varying the rate does not affect the performance too much. For the depth of the tree, we can see that if the tree goes too deep (i.e., the depth is greater than 5), it may have a serious overfit problem due to the excessive capability, and if the tree is too shallow (i.e., the depth is smaller than 5), it cannot provide enough capability enhancement for prediction. 

\subsubsection{Model Size Comparison}

In order to perform a fair comparison between the base
model (without using our framework) and our framework,
we conduct experiments so that they have the same model complexities. 
Specifically, after we obtain SR-PredAO(SGNN-HN), 
we enlarge the base model (i.e., SGNN-HN) by increasing the embedding dimensionality and this base model (without using our framework), after parameter-tuning, is regarded as a baseline. 
The experimental result on Diginetica is shown in Table \ref{tab:size}. The enlarged base model cannot outperform SR-PredAO(SGNN-HN) due to the inappropriate training capacity increment of the base model.

\begin{table}[h]
    \centering
    {\begin{tabular}{llll}
    \toprule
    \textbf{Model}                  & \textbf{Size} & \textbf{HR@20} & \textbf{MRR@20} \\
    \midrule
    Enlarged SGNN-HN        & \num{1605}MB & 55.24 & 18.64 \\
    \midrule    
    SR-PredAO(SGNN-HN)            & \num{1118}MB & \textbf{55.91} & \textbf{18.78} \\
    \bottomrule
    \end{tabular}}
    \caption{Size comparison (\%) result on Diginetica}
    \label{tab:size}
\end{table}

\subsection{Degrees-of-Freedom Study}\label{sec:dof}
This section gives a comprehensive and quantitative study of the capability of the NDF-SR module to show the wide bandwidth of capability our model can provide. As stated in Section \ref{prob def}, we use \textit{Degrees-of-Freedom} (DoF) to formally define the capability of a model in this paper. Following \cite{mentch2020randomization}, the DoF is a good illustration of model's capability. It is defined as follows: Assuming we have a dataset $\textit{D}_n = \{(\bm{x}_i, y_i)\}_{i = 1}^{M}$, where $\bm{x}_i \in \mathbb{R}^{n'}$ and $\varepsilon_i \overset{i.i.d.}{\sim} \mathcal{N}(0, \sigma^2)$. The relationship between $\bm{x}$ and $y$ is in the form $y_i = f(\bm{x}_i) + \varepsilon_i$, where $f$ is the true relation between $\bm{x}$ and $y$. The model we fit to estimate $f$ is denoted as $\hat{f}$, in our case, and prediction is $\hat{y}_i = \hat{f}(\bm{x}_i)$, it can be either the base model or \method enhanced model. Then, the DoF is defined as $DoF(\hat{f}) = \frac{1}{\sigma^2}\sum_{i = 1}^{N} \text{Cov}(\hat{y}_i, y_i)$. 

Since in the session-based recommendation problem, we do not know the true function $f$, we perform DoF analysis on two simulated underlying functions, namely the "MARSadd"\cite{mentch2020randomization}: $y_i = 0.1e^{4x_{i1}} + \frac{4}{1 = e^{-20(x_{i2} - 0.5)}} + 3x_{i3} + 2x_{i4} + x_{i5} + \varepsilon_i$, where $\epsilon \overset{i.i.d.}{\sim} \mathcal{N}(0, 1)$ and $\bm{x}_i = [x_{i1}, \cdots, x_{i5}]^\top$ are randomly sampled from uniform distribution between 0 and 1. Another underlying function is we proposed "POWERadd": $y_i = \sum_{j = 1}^5 x_{ij} + \sum_{j = 6}^{10} x_{ij} + \varepsilon_i$, this function in order to test the model's DoF behavior under high-order curvature and extra dimension. Data and error terms are sampled as "MARSadd".

The experimental results are demonstrated in Fig. \ref{fig:dof}. We can observe that except for extreme cases (like $r = 0.9$ or $depth = 1$), NDF-SR shows significantly higher DoF than the linear model. Additionally, by controlling depth and pruning rate (r), we can achieve a very flexible change in DoF in both experiments. This is further evidence of the effectiveness of pruning. From the results of simple simulated functions, we can easily extrapolate that the linear model suffers from a low-capability problem, while the NDF-SR (or DoF) we proposed can provide a higher and more controllable capability.

\subsection{Summary}
In summary, framework \emph{SR-PredAO}, when applied to existing state-of-the-art models, could have up to 2.9\% improvement on HR@20 and 2.3\% of improvement on MRR@20. We can observe this improvement in almost all base models on all datasets. By considering the consistency of improvement and the ease of applicability of our framework, we regard our contribution as a major improvement to the field of the session-based recommendation system. 

%% file: 6-conclusion.tex
\section{Conclusion}\label{conclusion}

In this paper, we are the first to discover the important low-capability issue in the predictor module of most (if not all) existing models, lowering down their prediction accuracy. 
To address this important issue, we propose a framework called \emph{SR-PredictAO} which could be applied
to any existing models following the common encoder-predictor paradigm. Extensive experimental results 
on two public benchmark datasets show that when framework \emph{SR-PredictAO} is applied to 3 existing state-of-the-art models, their performance are consistently improved up to 2.9\% on HR@20 and up to 2.1\% on MRR@20. Due to the consistent improvement on all datasets, we regard our contribution as a major improvement to the field of the session-based recommendation system. 

\section{Future Work}

Although SR-PredAO sets an effective and general enhancement that can be applied for all models that lack capability in modeling complex underlying behaviors. There are many future possible studies that can be developed based on SR-PredAO. Firstly, the cost of this framework is relatively high when the number of leaf nodes is large. Thus, how to develop an efficient tree-based method is a potential direction. Secondly, this paper only discusses the task in session-based recommendation, and the tree-based enhancement can be applied to a wider field such as language and vision modeling. Thirdly, the theoretical foundation for neural-based trees also needs to be built up by future studies.

\section{Acknowledgement}
We are grateful to the anonymous reviewers for their constructive comments. The research is supported in part by WEB24EG01-H.

%% file: 7-appenidx.tex
\appendix

\subsection{Proof of Alleviator}\label{JS-prove}

\subsubsection{Assumption 1}
All session data are i.i.d. (i.e., independent and identically distributed) samples affected by random user's behavior under some uniform distribution $\mathbb{P}_\textbf{s}$.  

i.e. For a set of collected sampled sessions: $\{\bm{s}_i\}_{i = 1}^m$, where $\bm{s}_i = [s_{i, 1}, s_{i, 2}, \cdots, s_{i, l_i}]$ as we defined in Section \ref{prob def}. For all $i$, we have: 
$\bm{s}_i - \epsilon_i \sim \mathbb{P}_s$
where $\epsilon_i$ is the random user's behavior, which can be intuitively understood as the user's random behavior

\subsubsection{Assumption 2 (Empirical Bayes Assumption I)}
The encoded value of a session (i.e., $z_{ij}$ in Section \ref{JS-Estimator}) is not the real value, just an observation affected by random user's behavior:

i.e., for a not affected session $\bm{s}_i - \epsilon_i$; the real encoded latent variable is 
$\bm{\mu}_i = [\mu_{i1}, \cdots, \mu_{in'}]^T
$
; For a fixed $j = 1, 2, \cdots, n'$ (where $n'$ is the dim for latent variable), the $\{\mu_{ij}\}_{i = 1}^m$ and follows distribution $\mathbb{P}_s^{(j)}$. 

That means the real value for $j-th$ propriety of the $i-th$ session should be $\mu_{ij}$; 
But due to the effect of random user's behavior, the encoded result we observe from the session-encoder is $z_{ij}$

\subsubsection{Assumption 3 (Empirical Bayes Assumption II)}
The observed value of the encoded session $z_{ij}$ follows the distribution of $\mathcal{N} (\mu_{ij}, \sigma_j^2)$ and $\sigma_j \geqslant 1$ (if this assumption is not met, we can always do batch normalization to make the $\sigma_j$ not far from 1).

The Normal distribution assumption comes from the statistic common that if a distribution is affected by extremely complex factors, like the random user's behavior. The safest way is to assume that they are normally distributed. Since all numbers in $\{z_{ij}\}_{i = 1}^m$ represent the same factor of the session (the $j-th$ encoded factor), it is reasonable to assume they have the same and relatively large variance. 

\subsubsection{Target}
In high-level understanding, what we observed in the real data is not the full fact but noisy data that have information of the underlying true value. Our goal is to obtain the underlying true value (i.e., $\mu_{ij}$ in our case) through observed values (the $z_{ij}$ in our case). 

The rigorous definition of the target is: given a batched, observed encoded result: 
$\bm{Z}\in \mathbb{R}^{m \times n'}$ and its corresponding underlying true value $\bm{\mu} \in \mathbb{R}^{m \times n'}$

For a fixed $j \in [1, n']$, get an estimator $\hat{\mu}_{ij} | \bm{\xi}_j$ for $\mu_{ij}$ s.t. $\mathbb{E} [\sum_{i = 1}^m (\hat{\mu}_{ij} - \mu_{ij})^2] $ is small. 

Consider the Max likelihood estimator $\hat{\mu}_{ij}^{(MLE)} = z_{ij}$, and $\hat{\mu}^{ij}_{(JS)} = (1 - \frac{m - 2}{\|\bm{\xi}_j\|^2}) z_{ij}$. 

Claim that: $\mathbb{E}[\sum_{i = 1}^m (\mu_{ij} - \hat{\mu}_{ij}^{(JS)})^2] \leqslant \mathbb{E} [\sum_{i = 1}^m (\mu_{ij} - \hat{\mu}_{ij}^{(MLE)})^2] $

\subsubsection{Proof of claim}

$
      \mathbb{E} [\sum_{i = 1}^m (\mu_{ij} - \hat{\mu}_{ij})^2] 
    = \sum_{i = 1}^m \mathbb{E} [(z_{ij} - \hat{\mu}_{ij})^2 - (z_{ij} - \mu_{ij})^2 + 2(\hat{\mu}_{ij} - \mu_{ij})(z_{ij} - \mu_{ij})]
    = \sum_{i = 1}^m \mathbb{E} [(z_{ij} - \hat{\mu}_{ij})^2] - m \cdot \sigma_j + 
      2\sum_{i = 1}^m \mathbb{E} [(\hat{\mu}_{ij}- \mu_{ij})(z_{ij} - \mu_{ij})]
$

Consider distribution function for $z_{ij}$ as: $\varphi(z_{ij} | \mu_{ij}, \sigma_j) = \frac{1}{\sqrt{2\pi} \cdot \sigma_j} 
\exp(-\frac{(z_{ij} - \mu_{ij})^2}{2 \sigma^2_j})$. Therefore, $(z_{ij} - \mu_{ij})(\hat{\mu}_{ij} - \mu_{ij}) = -\sigma_j^2 \frac{\partial}{\partial z_{ij}} \varphi(z_{ij} | \mu_{ij}, \sigma_j)$ 

Therefore, for any continuous, differentiable, and $|f(z)| < \infty$, function $f: \mathbb{R} \rightarrow \mathbb{R}$. For simplicity, denote $\varphi(z_{ij}|\mu_{ij}, \sigma_j)$ as $\varphi(z_{ij})$ we have:

\[\begin{aligned}
     & \mathbb{E} [(z_{ij} - \mu_{ij}) f(z_{ij})] \\
    =& \int_{-\infty}^{+\infty} (z_{ij} - \mu_{ij})f(z_{ij}) \varphi(z_{ij}) dz_{ij} \\
    =& (-\sigma_j^2) \cdot (\int_{-\infty}^{+\infty} (\frac{\partial}{\partial z_{ij}}\varphi(z_{ij}))f(z_{ij}) dz_{ij}) \\
    =& (-\sigma_j^2) \cdot \varphi(z_{ij}) \cdot f(z_{ij}) |^{+\infty}_{-\infty} - \int_{-\infty}^{+\infty} f'(z_{ij})\varphi(z_{ij})          dz_{ij} \\
    =& \sigma_j^2 \cdot (\int_{-\infty}^{+\infty} f'(z_{ij}) \varphi(z_{ij}) dz_{ij}) \\
    =& \sigma_j^2 \mathbb{E}[\frac{\partial}{\partial z_{ij}} f(z_{ij})]
\end{aligned}\]

Therefore, we have: $\mathbb{E}[(z_{ij} - \mu_{ij})(\hat{\mu}_{ij} - \mu_{ij})] = \mathbb{E} [\frac{\partial \hat{\mu}_{ij}}{\partial z_{ij}}] \cdot \sigma_j^2 $. Therefore, when $\hat{\mu}_{ij}$ is MLE: $\mathbb{E}[\sum_{i = 1}^m (\mu_{ij} - \hat{\mu}_{ij}^{(MLE)})^2] = 0 - m \cdot \sigma_j^2 + 2m \cdot \sigma_j^2 = m \cdot \sigma_j^2$
When $\hat{\mu}_{ij}$ is $\hat{\mu}_{ij}^{(JS)} $. We have: $\mathbb{E}[\frac{\partial \hat{\mu}_{ij}^{(JS)}}{\partial z_{ij}}] = 1 - \frac{m - 2}{\|\bm{\xi}_j\|^2} + \frac{2(m - 2) z_{ij}^2}{\|\bm{\xi}_j\|^4}$

Therefore, 
$\sum_{i = 1}^m \mathbb{E} [(\hat{\mu}_{ij}^{(JS)} - \mu_{ij})(z_{ij} - \mu_{ij})] = m - \mathbb{E} [\frac{(m - 2)^2}{\|\bm{\xi}_j\|^2}]$

With $\mathbb{E}[(z_{ij} - \hat{\mu}^{(JS)}_{ij})^2] = \mathbb{E}[\frac{(m - 2)^2}{\|\bm{\xi}_j\|^2} z_{ij}^2] $, we have: 
$\mathbb{E} [\sum_{i = 1}^m (\mu_{ij} - \hat{\mu}_{ij}^{(JS)})^2] = m \cdot \sigma_j^2 + m(1 - 2\sigma_j^2) \mathbb{E} [\frac{(m - 2)^2}{\|\bm{\xi}_j\|^2}]$

Since in our assumption $\sigma_j^2 \geqslant 1$, we have: 
\begin{equation}
    \mathbb{E}[\sum_{i = 1}^m (\mu_{ij} - \hat{\mu}_{ij}^{(JS)})^2] \leqslant \mathbb{E} [\sum_{i = 1}^m (\mu_{ij} - \hat{\mu}_{ij}^{(MLE)})^2]
\end{equation}